\documentclass[aps,prl,twocolumn]{revtex4-1}
\usepackage{amsmath}
\usepackage{amssymb}
\usepackage{graphicx}

\begin{document}

\title{Majorana's stellar representation for the local polarization \\ of 
	harmonic electromagnetic and gravitational waves}

\author{Patrick Bruno}
\email[]{patrick.bruno@esrf.fr}

\affiliation{European Synchrotron Radiation Facility, BP 220, F-38043 Grenoble Cedex 9, France}

\date{18 March 2019}

\begin{abstract}
The local polarization of electromagnetic (EMW) and gravitational waves (GW) is discussed from an operational point of view, in which all the relevant mathematical framework is \emph{constructed} in terms of measurements of the power absorbed by a local detector. The intrinsic dependence of the observations upon the nature of the detector is emphasized. In particular, the benefit of using a dual-symmetric detector, equally sensitive to the electric and magnetic fields of the EMW (resp. gravito-electric and gravito-magnetic tensor in the GW case) is pointed out. The Majorana stellar representation of the polarization is introduced, and its physical interpretation is highlighted. Finally, expressions for the energy density, linear momentum density, helicity and spin density of the wave in terms of the Majorana representation are presented.
\end{abstract}


\maketitle

For vectorial and rank-2 tensorial wave, polarization is a concept of central importance. The tensorial rank of the wave is directly related to the spin of the quantum particle associated with wave: $S=1$ for the photon, $S=2$ for the graviton. The aim of the present paper is to discuss the local polarization of electromagnetic (EMW) and gravitational waves (GW). This will be done by elaborating upon a \emph{gedankenexperiment} in which the wave is probed by a local intensity detector. I shall first point out how the necessary mathematical framework can be \emph{constructed} from physical measurements performed with the detector. Then I shall introduce the stellar representation first proposed by Majorana to describe the pure quantum states of a quantum spin system and highlight its physical significance in the present context. Finally, I shall indicate how physical quantities characterizing the wave can be expressed in terms of Majorana's stellar representation.

All waves considered here are assumed to be perfectly coherent harmonic waves of angular frequency $\omega$. In particular, I consider the situation in which one studies the wave by means of an \emph{intensity} detector $D$, i.e., by measuring the power absorbed by the detector as it is placed at a given (fixed) position in the wave-field. At this point, there is no need to specify whether one considers EMWs or GWs. The wave-fields, being solutions of a linear (or linearized) wave equation, naturally belong to an infinite-dimensional vector space $\mathcal{V}$. Let us now considered the more specific mathematical structure than can be constructed with the help of our detector. Let $A(\mathbf{r})$ be the complex amplitude (of the appropriate tensorial character) of our harmonic wave-field \cite{sign convention}. From physical considerations, it is obvious that the signal measured by the detector is a functional $I_D[A(\mathbf{r})]$ satisfying the following conditions:
%
$I_D[A(\mathbf{r})] \geq 0 $ and 
$I_D[ \alpha A(\mathbf{r})] =  |\alpha|^2 I_D[A(\mathbf{r})]  , \ \forall \alpha \in \mathbb{C} .$
%
However, depending on the detector used and its position in the wave-field, some non-zero wave-fields may yield a vanishing signal in the detector, for example if the latter happen to probe only nodal regions of the wave-field. The \emph{operational} viewpoint adopted here leads us to acknowledge this and "remove" from our consideration the waves for which our detector is blind. The is done in practice by introducing the following equivalence relation $\stackrel{D}{\sim}$:
%
$A \stackrel{D}{\sim} B \Leftrightarrow I_D[A-B] =0.$
%
Thus we say that two wave-fields $A$ and $B$, \emph{as seen by the detector} $D$, are equivalent \emph{iff} the detector "does not see" the difference between $A$ and $B$. The equivalence classes 
%
$A_D \stackrel{\text{def}}{=} \{B\in \mathcal{V}; B \stackrel{D}{\sim} A \} ,$
%
constitute a quotient set
%
$\mathcal{V}_D \stackrel{\text{def}}{=} \mathcal{V} / \stackrel{D}{\sim} .$
%
The quotient set $\mathcal{V}_D$ naturally inherits from $\mathcal{V}$ a vector space structure on its own, of dimension lower or equal to that of $\mathcal{V}$. 

Let us further assume that, when superposing two wave-fields $A$ and $B$, we have the possibility of dephasing one with respect to the other; this would be the case for instance if the two wave fields are derived from a common coherent source, and a delay line is inserted in one the branches of the interferometer \cite{gedanken}. Thus, we can study $I_D[A+e^{-i\phi}B]$ as a function of the phase shift $\phi$. Now, one can readily observe that for any conceivable intensity detector operated in the linear response regime, the intensity of a linear superposition of waves is of the form
\begin{eqnarray}
I_D[\alpha A+ \beta e^{-i\phi}B] &=& \alpha^2 I_D[A] + \beta^2 I_D[B] \\ 
&+& 2\alpha\beta\Delta_{AB} \cos(\phi -\phi_{AB} ) , 
\end{eqnarray}
$\forall\alpha,\beta\in \mathbb{R}$, (and in any case, we restrict the discussion to detectors satisfying this condition). This enables us to define
\begin{equation}
\langle A|B \rangle_D \stackrel{\text{def}}{=} \Delta_{AB} e^{i\phi_{AB}},
\end{equation}
which, as one can easily check, satisfies all axioms of an inner product, thereby endowing $\mathcal{V}_D$ with the structure of an inner product space. Further, the intensity measured by the detector is actually a norm on $\mathcal{V}_D$: $\| A\|_D \stackrel{\text{def}}{=} \sqrt{\langle A |A\rangle_D}$. 
Since for any physically realizable wave-field $A$ and detector $D$, the absorbed power is finite, i.e., $I_D[A] = \|A\|_D^2 < + \infty$, $\mathcal{V}_D$ actually acquires the structure of a Hilbert space \cite{Neumann2018}. It is important to emphasize that the mathematical structures we have thus constructed does not result from formal considerations, but rest entirely on the physical measurements we may carry out using our detector. Likewise, one should keep in mind that, given two detectors $D$ and $D^\prime$, one would obtain two distinct Hilbert spaces $\mathcal{V}_D$ and $\mathcal{V}_{D^\prime}$, endowed with two distinct inner products $\langle A|B \rangle_D$ and $\langle A|B \rangle_{D^\prime}$. In other words, one is confronted with an essential \emph{detector subjectivity}, and must take it into account explicitly when discussing \emph{objectively} any outcome of our experiments.

Since we are interested primarily in the \emph{polarization} of the wave-field, the mere amplitude and phase of the waves are essentially of no interest, and we may want to discard them from our consideration. Since the inner product $\langle A|B \rangle_D$ changes as we change the amplitude and/or phase of $A$ and/or $B$ without changing their polarization, this is clearly not the appropriate tool to compare the polarizations of waves $A$ and $B$. Thus I proceed to consider as equivalent waves that differ merely in their amplitude and/or phase, i.e., for $A$ and $B$ with  $\| A\|_D \neq 0$ and $\| B\|_D \neq 0$, I define
%
$A_D \stackrel{D}{\equiv} B_D \Leftrightarrow \forall C_D \in \mathcal{V}_D , \frac{|\langle A|C \rangle_D|}{\| A\|_D} = \frac{|\langle B|C \rangle_D|}{\| B\|_D}.$
%
This is an equivalence relation, and the corresponding quotient set is the projective Hilbert space  $\mathcal{PV}_D \stackrel{\text{def}}{=} \mathcal{V}_D / \stackrel{D}{\equiv}$, which our task is to characterize. The projective Hilbert space is not a vector space; rather it is a complex manifold, wich can be equipped with various structures, in particular with a metric and with a symplectic structure. Here, we are specifically interested with the metric structure, since a metric would enable us to express quantitatively the closeness or remoteness of the polarizations of wave-fields. Following the \emph{operational} viewpoint adopted here, we want to do this on the basis of experimental information obtained via our detector. We thus introduce the Fubini-Study distance $d_{AB}$:
%
$\cos d_{AB} \stackrel{\text{def}}{=} \frac{|\langle A | B\rangle|}{\|A\|\, \|B\|}, \ \text{with}\ 0 \leq d_{AB} \leq \frac{\pi}{2}.$
%
Here an in the remaining of this paper, unless necessary, we lighten the notations by removing the detector label $D$. The maximum distance between waves in $\mathcal{PV}_D$ is $\pi/2$. A pair of waves for which the interference intensity vanishes are maximally distant, i.e., \emph{antipodal} to each other.

In order to proceed further, it is necessary to be more specific about the waves and detectors considered. Since our goal is to address the \emph{local} polarization of the wave, we choose a spherically symmetrical, isotropic detector of dimensions small with respect to the wavelength. For an EMW, we may for example consider a small metal sphere. The intensity measured will be of the form (up to some unimportant prefactor)
\begin{equation}
I_{\text{EM},w} = (1-w)\ \mathbf{E}\cdot \mathbf{E}^\star + w\ \mathbf{B}\cdot \mathbf{B}^\star ,
\end{equation}
where $\mathbf{E}$ (resp. $\mathbf{B}$) is the complex amplitude of the electric (resp. magnetic) field (expressed here in Gaussian units) at the center of the detector, and is characterized by the electric-magnetic (e-m) weight parameter $w$. The electric and magnetic dipolar polarizabilities can be calculated from Mie theory. The ratio of the magnetic and electric absorption coefficients, for sufficiently long wavelength is
%
$\frac{w}{1-w}=\frac{\mathrm{Im}(\alpha_B)}{\mathrm{Im}(\alpha_E)} = \frac{8\pi^2}{45} \left(\frac{R \sigma}{c}\right)^2,$
%
where $R$ is the radius, $\sigma$ the electric conductivity, and $c$ the velocity of light \cite{Asenjo2012}. Thus, depending of the value of these parameters, our detector will be sensitive to the electric ($w=0$) or magnetic ($w=1$) intensity, or to a weighted average of them ($0<w<1$).  

In close analogy with EMWs, GWs comprise gravito-electric (GE) and gravito-magnetic (GM) components \cite{Owen2011,Nichols2011,Barnett2014,footnote-GE-GM}, the complex amplitude of which, for harmonic radiation of angular frequency $\omega$, is given, respectively, by the traceless symmetric (complex) tensors
\begin{equation}
E_{ij} = \frac{\omega^2}{2}\  h_{ij},\ \text{and} \
B_{ij}=\frac{-i\omega c}{2}  \ \epsilon_{ikl}\ \partial_k h_{lj} ,
\end{equation}
where, $\epsilon_{ikl}$ is the Levi-Civita antisymmetric tensor, and Einstein summation convention over repeated dummy indices has been used. The symmetric traceless tensor $h_{ij}$ is the (complex) amplitude of the deviation of the metric tensor with respect to flat Minkowski spacetime. The GE field $E_{ij}$ is the \emph{tidal field} tensor \cite{Owen2011,Nichols2011}; its effect on a spherical resonator, for example, is to induce a spheroidal quadrupolar stress \cite{Maggiore2008}. The GW power absorbed by a spherical mass resonator of radius small with respect to the wavelength is (up to a for us unimportant prefactor) $I_{\text{GW},e} = E_{ij}E_{ij}^\star$ \cite{Maggiore2008}. 
 
The GM tensor $B_{ij}$ is the \emph{frame-drag} field, which causes nearby free-falling gyroscopes to precess with respect to each other \cite{Owen2011,Nichols2011}. It is considered to have no effect on mass-resonators \cite{Maggiore2008}, a fact which can be traced back to the absence of a GM analog of Faraday's induction \cite{Costa2007,Costa2008,Costa2016}.  It can be detected, however, by means of suitably designed Sagnac interferometry \cite{Frauendiener2018a,Frauendiener2018b}. Without specifying a particular design, we assume that one can measure the GM intensity $I_{\text{GW},b} = B_{ij}B_{ij}^\star$, and we consider a detector measuring a weighted average of the GE and GM intensities, i.e.,
\begin{equation}
I_{\text{GW},w} = (1-w)\ E_{ij}E_{ij}^\star + w\ B_{ij}B_{ij}^\star .
\end{equation}
 
For an EMW (resp. GW) \emph{propagative} \emph{plane wave}, the intensity measured is independent of the e-m weight parameter $w$, since $\mathbf{E}\cdot \mathbf{E}^\star = \mathbf{B}\cdot \mathbf{B}^\star$ (resp. $E_{ij}E_{ij}^\star = B_{ij}B_{ij}^\star$) in that case. However, this no longer holds for a \emph{significantly non-planar} EMW or GW, a fact that has been recognized long ago in the EMW case \cite{Nye1987,Berry2001}, so that detectors of different e-m weight parameter $w$ will generally measure different intensities.  

Let us consider first the case of an electric or GE detector ($D= "e"$, i.e., $w=0$). Since, the detector is blind to the magnetic component of the radiation ($\mathbf{B}$, or $B_{ij}$), the dimension of our Hilbert space $\mathcal{V}_e$ is 3 (resp. 5) for EMW (resp. GW) radiation, corresponding to the 3 (resp. 5) complex numbers needed to specify the electric field $\mathbf{E}$ (resp. the traceless symmetrical GE tensor $E_{ij}$). Thus our problem can be mapped onto that of a quantum spin $S=1$ (resp. $S=2$), and the determination of the polarization state of the wave is equivalent to that of the quantum state of the spin. The quantum state is characterized in the projective Hilbert space $\mathcal{PV}_e$, by 4 (resp. 8) real coordinates. The mapping onto the quantum spin problem is such that left-circularly polarized (LCP) plane-waves (PW) \cite{phase definition} $C^L_{\hat{\mathbf{n}}}$ of direction $\hat{\mathbf{n}}$ are mapped onto the spin coherent states $|\hat{\mathbf{n}}\rangle$ \cite{Radcliffe1971,Arecchi1972}, i.e., the eigenstates of $\hat{\mathbf{n}}\cdot\mathbf{S}$ with eigenvalue 1 (resp. 2). With the help of our electric intensity detector, we can measure the Bargmann function of wave $A$:
\begin{equation}
A^L(\hat{\mathbf{n}}) \stackrel{\text{def}}{=} \langle C^L_{\hat{\mathbf{n}}} | A \rangle = \left\lbrace 
\begin{array}{l}
\mathbf{e}_{\hat{\mathbf{n}}}^{L\star} \cdot \mathbf{E},\ \text{(EMW case)}, \\
e_{\hat{\mathbf{n}},ij}^{L\star} \, E_{ij}\ \text{(GW case)},
\end{array} \right.
\end{equation}
where $\mathbf{e}^L_{\hat{\mathbf{n}}}$ (resp. $e^L_{\hat{\mathbf{n}},ij}$) is the complex electric field (resp. GE tidal tensor) of the LCP-PW $C^L_{\hat{\mathbf{n}}}$, as well as the corresponding Husimi function
%
$Q^L_A(\hat{\mathbf{n}}) \stackrel{\text{def}}{=} |A^L(\hat{\mathbf{n}})|^2 ,$
%
which expresses the distance $d_{A,L\hat{\mathbf{n}}}$ between waves $A$ and $C^L_{\hat{\mathbf{n}}}$:
%
$\cos^2 d_{A,L\hat{\mathbf{n}}} = \frac{Q^L_A(\hat{\mathbf{n}})}{I_e(A)\ I_e(C^L_{\hat{\mathbf{n}}})} .$
%
The Bargmann function $A^L(\hat{\mathbf{n}})$ can be thought of as a wavefunction on the unit sphere, that entirely characterizes the wave $A$, with corresponding probability density given by the Husimi function $Q^L_A(\hat{\mathbf{n}})$. 

The usefulness of the Bargmann and Husimi functions stems from an important property discovered by Majorana \cite{Majorana1932} in the context of quantum spin systems, which in the context of our problem, says: for a given EMW (resp. GW) $A$, the Bargmann and Husimi functions possess exactly 2 (resp. 4) (non-necessarily distinct) zeros on the unit sphere, called Majorana stars (MSs), and the constellation of the (undiscernable) MSs uniquely determines the electric polarization of the wave. In other words, there are 2 (resp. 4) LCP-PW that are antipodal to $A$. Majorana's representation of quantum spin systems is discussed in detail in Ref.~\cite{Bruno2012}, whereas the proof of Majorana's theorem in the present context is presented in the appended Supplemental Materials. One notices that a constellation of 2 (resp. 4) stars is specified by 4 (resp. 8) angle coordinates which coincides with the (real) dimension of the projective Hilbert space for the EMW (resp. GW) case, as it should. Thus, the knowledge of the Majorana constellation (MC), together with the value of the Bargmann function at any point where it does not vanish, fully specify the electric field $\mathbf{E}$ (resp. GE tidal tensor $E_{ij}$).

Let us now illustrate Majorana's stellar representation by discussing the MCs obtained for archetypal waves. We note in passing that, if we would have performed the same construction with right-circularly polarized (RCP) PW, we would have obtained a MC consisting of the antipodes of the MC obtained from interference with LCP-PWs \cite{footnote-LCP-RCP}. For EMW, a LCP (resp. RCP) PW propagating along the $+\hat{\mathbf{z}}$ axis has its two MSs collapsed at the south (resp. north) pole of the unit sphere, whereas a PW linearly polarized along the $x$ axis has its two MSs located on the equator along the $+\hat{\mathbf{x}}$ and $-\hat{\mathbf{x}}$ axes, irrespective of its propagation direction in the $yz$ plane. An elliptically polarized PW has an intermediate MC; an interesting geometrical construction of the MC in terms of the polarization ellipse has been proposed \cite{Hannay1998}. An alternative construction is as follows: considering the polarization ellipse of wave $A$, the two MSs are given by the two LCP-PWs for which the polarization circle, projected on the plane of the ellipse of $A$, yields and ellipse rotated by $\pi /2$ with respect to the ellipse of wave $A$, with opposite circulation direction \cite{footnote-ellipse}. Note that nothing in the (electric) MC of an EMW enables to distinguish whether the wave is or not a propagative PW.

Turning to the GW case with $w=0$ (gravito-electric detector), a LCP (resp. RCP) gravitational PW propagating along the $+\hat{\mathbf{z}}$ axis has its two MSs collapsed at the south (resp. north) pole of the unit sphere, whereas a linearly polarized PW with {\large \sf X} (resp. {\large \bf +}) polarization has its MSs located on the (resp. rotated by an angle $\pi/4$ from the) $+\hat{\mathbf{x}}$, $-\hat{\mathbf{x}}$, $+\hat{\mathbf{y}}$ and $-\hat{\mathbf{y}}$ directions. Elliptically polarized waves have an intermediate MC, with the MSs moving symmetrically towards the $+\hat{\mathbf{z}}$ as the polarization progressively evolves from linear to RCP. A important observation is that for a gravitational PW the MC has a four-fold symmetry around the propagation direction; however, this is generically not the case for a non-planar-propagative wave.Thus any deviation from perfect four-fold symmetry of the MC is a signature of the non-planar-propagative character of the wave, a feature not available for EMW. Conversely, however, a GW with four-fold-symmetric MC is not necessarily a propagative PW.

If one replaces the electric intensity detector by magnetic intensity detector ($w=1$), one then obtains similar information about the polarization of the magnetic part of the wave ($\mathbf{B}$, resp. $B_{ij}$). For propagative PWs, the electric and magnetic part of the wave are simply related to each other: the magnetic MC of a EMW (resp. GW) propagative plane-wave is obtained from the electric MC by a rotation of $\pi/2$ (resp. $\pi/4$) around the propagation axis. This, however, is not the case for a generic wave, and, as we shall see, a more complex, but richer, situation arises if we use a detector that is sensitive to both the electric and magnetic component of the wave, i.e., for $0<w<1$. The first point to observe is that, since the detector is sensitive to both the electric and magnetic components of the radiation, the dimension of the Hilbert space is now twice as big: for an EMW (resp. GW) one needs 6 (resp. 10) complex numbers to fully describe the electric and magnetic components of the wave $\mathbf{E}$ and $\mathbf{B}$ (resp. $E_{ij}$ and $B_{ij}$). Our problem can be mapped onto a system with $S=1$ (resp. $S=2$) with 2 available sites $M_e$ and $M_b$. The spins $S_e =1$ (resp. $S_e=2$) on site $M_e$ and $S_b =1$ (resp. $S_b=2$) on site $M_b$ describe, respectively, the polarizations of the electric and magnetic components of the wave, whereas relative amplitude of the wave function encodes the relative magnitude and phase of the electric and magnetic components of the wave. If we proceed as above, and build a Bargmann function with LCP-PW, we obtain
\begin{eqnarray}
A^L(\hat{\mathbf{n}})\! &\stackrel{\text{def}}{=}& \langle C^L_{\hat{\mathbf{n}}} | A \rangle \\
&=&\! \left\lbrace 
\begin{array}{l}
\!\!(1-w)\,\mathbf{e}_{\hat{\mathbf{n}}}^{L\star} \cdot \mathbf{E} +w\, \mathbf{b}_{\hat{\mathbf{n}}}^{L\star} \cdot \mathbf{B},\ \text{(EMW)}, \\
\!\! (1-w)\,e_{\hat{\mathbf{n}},ij}^{L\star} \, E_{ij}+w\,b_{\hat{\mathbf{n}},ij}^{L\star} \, B_{ij}\ \text{(GW)},
\end{array} \right. \\
&=&\! \left\lbrace 
\begin{array}{l}
\!\!\mathbf{e}_{\hat{\mathbf{n}}}^{L\star} \cdot \left[ (1-w)\,\mathbf{E} +i\,w\, \mathbf{B} \right],\ \text{(EMW)}, \\
\!\! e_{\hat{\mathbf{n}},ij}^{L\star} \left[ (1-w)\,\, E_{ij}+i\,w\, B_{ij}\right]\ \text{(GW)},
\end{array} \right.
\end{eqnarray}
since $\mathbf{b}^L_{\hat{\mathbf{n}}}=-i\, \mathbf{e}^L_{\hat{\mathbf{n}}}$ (resp. $b^L_{\hat{\mathbf{n}},ij}=-i\, e^L_{\hat{\mathbf{n}},ij}$). Thus, by using Majorana's procedure, one would obtain a MC that characterizes neither the electric nor the magnetic component of the wave, but some linear combination of them, to which we attribute no particular physical meaning. A solution to this problem is to use circularly polarized \emph{standing waves}, built by superposing counter-propagating LCP- and RCP-PWs: 
\begin{eqnarray}
C^e_{\hat{\mathbf{n}}} &\stackrel{\text{def}}{=}& \frac{1}{2}\left(C^L_{\hat{\mathbf{n}}} + C^R_{-\hat{\mathbf{n}}}\right), \\
C^b_{\hat{\mathbf{n}}} &\stackrel{\text{def}}{=}& \frac{1}{2i}\left(C^L_{\hat{\mathbf{n}}} - C^R_{-\hat{\mathbf{n}}}\right);
\end{eqnarray}
The former (resp. latter) is a purely (at the site of the detector) electric (resp. magnetic) circularly polarized standing wave. If one measures the 2 Bargmann functions
\begin{eqnarray}
A^e(\hat{\mathbf{n}}) &\stackrel{\text{def}}{=}& \frac{\langle C^e_{\hat{\mathbf{n}}} | A \rangle}{(1-w)} = \left\lbrace 
\begin{array}{l}
\mathbf{e}_{\hat{\mathbf{n}}}^{L\star} \cdot \mathbf{E},\ \text{(EMW case)}, \\
e_{\hat{\mathbf{n}},ij}^{L\star} \, E_{ij}\ \text{(GW case)},
\end{array} \right. \\
A^b(\hat{\mathbf{n}}) &\stackrel{\text{def}}{=}& \frac{\langle C^b_{\hat{\mathbf{n}}} | A \rangle}{w} = \left\lbrace 
\begin{array}{l}
i\,\mathbf{e}_{\hat{\mathbf{n}}}^{L\star} \cdot \mathbf{B},\ \text{(EMW case)}, \\
i\,e_{\hat{\mathbf{n}},ij}^{L\star} \, B_{ij}\ \text{(GW case)},
\end{array} \right. ,
\end{eqnarray}
one can apply Majorana's theorem and thus obtain the two MCs describing the polarizations of the electric and magnetic components of the wave. The relative magnitude and phase of the electric and magnetic components is obtained from the ratio $-i\, A^b(\hat{\mathbf{n}})/A^e(\hat{\mathbf{n}})$ for some vector $\hat{\mathbf{n}}$ that belongs to none of the two MCs. This procedure thus enables to determine both $\mathbf{E}$ and $\mathbf{B}$ (resp. $E_{ij}$ and $B_{ij}$), which fully characterizes the wave-field at the site of our detector. From those one can determine a number of physically relevant local (time-averaged) quantities characterizing the wave, which are listed in the appended Supplemental Materials.

The case of a \emph{dual-symmetric} detector ($w=1/2$) offers the particularly interesting possibility of analyzing \emph{directly} the two helicity components of the wave-field. Instead of splitting the wave-field in terms of its electric and magnetic components, as done above (which is meaningful from the point of view of the interaction of the wave with matter), one can analyze its two helicity components, by considering the Riemann-Silberstein vectors (resp. tensors) \cite{Aiello2015}
\begin{eqnarray}
\mathbf{R}^\pm &\stackrel{\text{def}}{=}& \frac{1}{2}\left( \mathbf{E} \pm i\, \mathbf{B} \right)\ \text{(EMW case)}, \\
R_{ij}^\pm &\stackrel{\text{def}}{=}& \frac{1}{2}\left( E_{ij} \pm i\, B_{ij}\right) \ \text{(GW case)} 
\end{eqnarray}
where the $\pm$ superscript labels the helicity of the wave; the usefulness of the helicity decomposition stems from the fact that, in vacuum, the $\pm$ components are decoupled from each other (unlike the electric and magnetic components which are intertwined by the wave equation). Therefore, the helicity decomposition may be considered as more fundamental, than the electric-magnetic decomposition, as far as the free field is concerned. If we do so, we can think of our problem as mapped onto a two-site spin system, corresponding to the 2 two helicity components ($\pm$) of the wave. Measuring the LCP-PW Bargmann function $A^L(\hat{\mathbf{n}})$ (see Eq.) directly enables to obtain the MC of the $(-)$-helicity component $\mathbf{R}^-$ (resp. $R_{ij}^-$) of an EMW (resp. GW), whereas a similar measurement of the RCP-PW Bargmann function $A^R(\hat{\mathbf{n}})$, yields the $(+)$-helicity component of the wave.

The helicity-analyzed physical quantities characterizing the wave are listed in the appended Supplementary Materials, together with their expressions in terms of the Majorana representation.

\begin{acknowledgments}
	
\end{acknowledgments}

\setcounter{equation}{0}
\renewcommand{\theequation}{S\arabic{equation}}

\onecolumngrid

\clearpage

\begin{center}
{\large \bf Supplemental Material for \\
	\textquotedblleft Majorana's stellar representation for the local polarization \\ of  
	harmonic electromagnetic and gravitational waves\textquotedblright }
\end{center}

\begin{center}
{Patrick Bruno (*)}\\
{\emph{European Synchrotron Radiation Facility, BP 220, F-38043 Grenoble Cedex 9, France}}\\
{18 March 2019}

\end{center}

\subsection{Proof of Majorana's theorem for electromagnetic and gravitational waves}

We look for the zero of the Bargmann function, which is
\begin{equation}
A^L(\hat{\mathbf{n}}) \stackrel{\text{def}}{=} \langle C^L_{\hat{\mathbf{n}}} | A \rangle = \left\lbrace 
\begin{array}{l}
\mathbf{e}_{\hat{\mathbf{n}}}^{L\star} \cdot \mathbf{E},\ \text{(EMW case)}, \\
e_{\hat{\mathbf{n}},ij}^{L\star} \, E_{ij}\ \text{(GW case)},
\end{array} \right.
\end{equation}
for the case of an electric-type detector ($w=0$). The other Bargmann functions used in the paper are similar and simply amount to replace the electric field (resp. tensor) by a linear combination of the electric and magnetic fields (resp. tensors).

For the EMW case, the complex electric field of a generic wave $A$ may be expressed as
\begin{equation}
\mathbf{E}=\sum_{m=-1}^{1}a_{1,m} Y_{1,m},
\end{equation}
with
\begin{eqnarray}
Y_{1,\pm1}&\stackrel{\text{def}}{=}&\frac{1}{\sqrt{2}}\left(
\begin{array}{c}
1 \\
\pm i \\
0
\end{array} \right) , \\
Y_{1,0}&\stackrel{\text{def}}{=}&\left(
\begin{array}{c}
0 \\
0 \\
1
\end{array} \right) .
\end{eqnarray}
For the GW case, we expand the complex GE tensor as
\begin{equation}
E=\sum_{m=-2}^{2}a_{2,m} Y_{2,m},
\end{equation}
with
\begin{eqnarray}
Y_{2,\pm 2}&\stackrel{\text{def}}{=}&\frac{1}{2}\left(
\begin{array}{ccc}
1 & \pm i & 0\\
\pm i & -1 & 0\\
0 & 0 & 0
\end{array} \right) , \\
Y_{2,\pm 1}&\stackrel{\text{def}}{=}&\frac{1}{2}\left(
\begin{array}{ccc}
0 & 0 & 1 \\
0 & 0 & \pm i \\
1 & \pm i & 0
\end{array} \right) ,\\
Y_{2,0}&\stackrel{\text{def}}{=}&\frac{1}{\sqrt{6}}\left(
\begin{array}{ccc}
1 & \pm i & 0\\
\pm i & -1 & 0\\
0 & 0 & 0
\end{array} \right) .
\end{eqnarray}

The electric field (resp. tensor) of the LCP-PW is 
\begin{eqnarray}
\mathbf{e}_{\hat{\mathbf{z}}}^{L} &=& Y_{1,1} \ \text{(EMW case)}, \\
e_{\hat{\mathbf{z}}}^{L} &=& Y_{2,2} \ \text{(EMW case)} ,
\end{eqnarray}
for $\hat{\mathbf{n}}=\hat{\mathbf{z}}$, and is obtained for $\hat{\mathbf{n}}$ pointing in some generic direction of polar angles $(\theta,\phi)$, by rotating $\mathbf{e}_{\hat{\mathbf{z}}}^{L}$ (resp. $e_{\hat{\mathbf{z}}}^{L}$) by angle $\theta$ around the $y$-axis, followed by a rotation of angle $\phi$ around the $z$-axis. This yields
\begin{eqnarray}
\langle C^L_{\hat{\mathbf{n}}} | Y_{1,1} \rangle &=& \cos^2 \left(\frac{\theta}{2}\right) \\
\langle C^L_{\hat{\mathbf{n}}} | Y_{1,0} \rangle &=& \sqrt{2}\cos \left(\frac{\theta}{2}\right) \sin \left(\frac{\theta}{2}\right) e^{-i\phi} \\
\langle C^L_{\hat{\mathbf{n}}} | Y_{1,-1} \rangle &=& \sin^2 \left(\frac{\theta}{2}\right) e^{-2i\phi} ,
\end{eqnarray}
for the EMW case, and
\begin{eqnarray}
\langle C^L_{\hat{\mathbf{n}}} | Y_{2,2} \rangle &=& \cos^4 \left(\frac{\theta}{2}\right) \\
\langle C^L_{\hat{\mathbf{n}}} | Y_{2,1} \rangle &=& 2\cos^3 \left(\frac{\theta}{2}\right) \sin \left(\frac{\theta}{2}\right) e^{-i\phi} \\
\langle C^L_{\hat{\mathbf{n}}} | Y_{2,0} \rangle &=& \sqrt{6}\cos^2 \left(\frac{\theta}{2}\right) \sin^2 \left(\frac{\theta}{2}\right) e^{-2i\phi} \\
\langle C^L_{\hat{\mathbf{n}}} | Y_{2,-1} \rangle &=& 2\cos \left(\frac{\theta}{2}\right) \sin^3 \left(\frac{\theta}{2}\right) e^{-3i\phi} \\
\langle C^L_{\hat{\mathbf{n}}} | Y_{2,-2} \rangle &=& \sin^4 \left(\frac{\theta}{2}\right) e^{-4i\phi} ,
\end{eqnarray}
for the GW case. Expressing $\hat{\mathbf{n}}$ in terms of its stereographic projection from the south pole on the (complex) equatorial plane $\hat{\mathbf{n}} \to z=e^{i\phi} \tan (\frac{\theta}{2})$, 
one gets
\begin{equation}
A^L(\hat{\mathbf{n}}) = \left\lbrace 
\begin{array}{l}
\frac{1}{1+|z|^2} \left( a_{1,1} + \sqrt{2} a_{1,0} z^\star + a_{1,-1} {z^{\star}}^2\right)  \ \text{(EMW case)}, \\
\frac{1}{(1+|z|^2)^2} (a_{2,2} + 2 a_{2,1} z^\star + \sqrt{6} a_{2,0} {z^\star}^2 + 2 a_{2,-1} {z^\star}^3 + a_{2,-2} {z^\star}^4 ) \ \text{(GW case)} .
\end{array} \right.
\end{equation}
Thus, the equation $A^L(\hat{\mathbf{n}}) =0$ is a polynomial equation in $z^\star$ of degree 2 (resp. 4) for the EPW (resp. GW) case, and Majorana's results follows immediately from the fundamental theorem of algebra. Note that if some degree of the polynomial happens to be lower than 2 (resp. 4), due to the vanishing of some of the coefficients $a_{l,m}$, then an according number of roots are located at infinity, corresponding to the south pole of the unit sphere.

The parallel with the physics of a quantum spin $S=1$ (resp. $S=2$) is obvious from the above, and is of course completely expected.

It is important to note that the above results holds because the LCP-PWs correspond to maximally polarized coherent states; it would not have been possible to obtain a Majorana-like representation from a Bargmann function constructed from linearly polarized states, as the Bargmann function would not have been an analytical function of $z^\star$, but rather would have involved both $z$ and $z^\star$.

\subsection{Expression of the wave local physical characteristics, in terms of the complex electric and magnetic wave components}

I list below the expressions of the local characteristics of the harmonic electromagnetic (EMW) (resp. gravitational wave (GW)), expressed in terms of the complex electric and magnetic components $\mathbf{E}$ and $\mathbf{B}$ (resp. $E_{ij}$ and $B_{ij}$), that have been recently derived from a dual-symmetric theory of electromagnetic  \cite{Cameron2012,Bliokh2013,Bliokh2014} and gravitational radiation \cite{Barnett2014}. The velocity of light $c$ and gravitational constant $G$ have been restored in the expressions below. 

Thus, we have the (time-averaged) energy density
\begin{eqnarray}
W_{\text{EMW}} &=& \frac{1}{16\pi}\left( \mathbf{E}^\star\cdot\mathbf{E}  + \mathbf{B}^\star\cdot\mathbf{B}  \right), \\
W_{\text{GW}} &=& \frac{c^2}{128\pi G\omega^2}\left( E_{ij}^\star E_{ij} + B_{ij}^\star B_{ij} \right) ,
\end{eqnarray}
the linear momentum density ($=$ energy flux density)
\begin{eqnarray}
\mathbf{P}_{\text{EMW}} &=& \frac{1}{8\pi c} \text{Re}\left( \mathbf{E}^\star\times\mathbf{B}  \right), \\
P_{\text{GW}\,i} &=& \frac{c}{64\pi G \omega^2}\epsilon_{ijk}\text{Re}\left( E_{jl}^\star B_{kl}  \right) 
\end{eqnarray}
the helicity density
\begin{eqnarray}
H_{\text{EMW}} &=& \frac{-1}{8\pi c\omega}\text{Im}\left( \mathbf{E}^\star\cdot\mathbf{B}   \right), \\
H_{\text{GW}} &=& \frac{-c}{32\pi G \omega^3}\text{Im}\left( E_{ij}^\star B_{ij} \right) ,
\end{eqnarray}
and the spin density ($=$ helicity flux density)
\begin{eqnarray}
\mathbf{S}_{\text{EMW}} &=& \frac{1}{16\pi\omega} \text{Im}\left( \mathbf{E}^\star\times\mathbf{E}  + \mathbf{B}^\star\times\mathbf{B}  \right), \\
S_{\text{GW}\,i} &=& \frac{c^2}{64\pi G\omega^3}\epsilon_{ijk}\text{Im}\left( E_{jl}^\star E_{kl} + B_{jl}^\star B_{kl} \right) .
\end{eqnarray}
We not that, while some quantities, such as the energy density and spin-density, can be analyzed as a sum of an electric and an magnetic contribution, some other, such as the linear momentum density and the helicity density are expressed as a product of the electric and magnetic fields (resp. tensors). The is ultimately due to the fact that the electric and magnetic components of the wave are intertwined by the wave equation.

Other quantities, such as the orbital and total angular momentum density, the spin and orbital contributions to the linear momentum density cannot be expressed solely in terms of the local values of the electric and magnetic field (resp. tensor), as they involve spacial derivatives of the fields (resp. tensors). They are therefore intrinsically not accessible from the kind of \emph{local} measurement discussed here (unless we move the detector).

\subsection{Expression of the wave local physical characteristics, in terms of the complex ($\pm$)-helicity components}

If, instead of the magnetic and magnetic field (resp. tensor), the above quantities are expressed in terms of the ($\pm$)-helicity components
\begin{eqnarray}
\mathbf{R}^\pm &\stackrel{\text{def}}{=}& \frac{1}{2}\left( \mathbf{E} \pm i\, \mathbf{B} \right)\ \text{(EMW case)}, \\
R_{ij}^\pm &\stackrel{\text{def}}{=}& \frac{1}{2}\left( E_{ij} \pm i\, B_{ij}\right) \ \text{(GW case)} 
\end{eqnarray}
as advocated in Ref.~\cite{Aiello2015}, one obtains the following expressions, respectively, for the energy density
\begin{eqnarray}
W_{\text{EMW}} &=& \frac{1}{32\pi}\left( \mathbf{R}^{+\star}\cdot\mathbf{R}^+  + \mathbf{R}^{-\star}\cdot\mathbf{R}^-  \right), \\
W_{\text{GW}} &=& \frac{c^2}{256\pi G\omega^2}\left( R_{ij}^{+\star} R_{ij}^+ + R_{ij}^{-\star} R_{ij}^- \right) , 
\end{eqnarray}
the linear momentum density (energy flux density)
\begin{eqnarray}
\mathbf{P}_{\text{EMW}} &=& \frac{1}{32\pi c} \text{Im}\left( \mathbf{R}^{+\star}\times\mathbf{R}^+ - \mathbf{R}^{-\star}\times\mathbf{R}^-  \right), \\
P_{\text{GW}\,i} &=& \frac{c}{256\pi G \omega^2}\epsilon_{ijk}\text{Im}\left( R_{jl}^{+\star} R_{kl}^+ - R_{jl}^{-\star} R_{kl}^- \right)  ,
\end{eqnarray}
the helicity density
\begin{eqnarray}
H_{\text{EMW}} &=& \frac{1}{32\pi c\omega}\left( \mathbf{R}^{+\star}\cdot\mathbf{R}^+  - \mathbf{R}^{-\star}\cdot\mathbf{R}^-    \right), \\
H_{\text{GW}} &=& \frac{c}{128\pi G \omega^3}\left( R_{ij}^{+\star} R_{ij}^+ - R_{ij}^{-\star} R_{ij}^- \right) ,
\end{eqnarray}
and the spin density (helicity flux density)
\begin{eqnarray}
\mathbf{S}_{\text{EMW}} &=& \frac{1}{32\pi\omega} \text{Im}\left( \mathbf{R}^{+\star}\times\mathbf{R}^+  + \mathbf{R}^{-\star}\times\mathbf{R}^-  \right), \\
S_{\text{GW}\,i} &=& \frac{c^2}{128\pi G\omega^3}\epsilon_{ijk}\text{Im}\left( R_{jl}^{+\star} R_{kl}^+ + R_{jl}^{-\star} R_{kl}^- \right) .
\end{eqnarray}
One notes that, as expected from the fact that the tow helicity components are decoupled from each other in the wave equation, all these quantities are naturally split as a sum of two pure-helicity terms, ith no cross-helicity terms. As pointed out in \cite{Aiello2015}, this is true only if a dual-symmetric formulation is employed.

An interesting point about the Majorana representation is that all the above physically relevant quantities can be expressed directly in terms the Majorana constellation, without even needing to obtain the polarizations vectors (resp. tensors) $\mathbf{R}^\pm$ (resp. $R_{ij}^\pm$) themselves, as shown in Ref.~\cite{Bruno2012}. I give below the results for EMW case (those for the GW are more lengthy, but can obtained straightforwardly by using the results derived in Ref.~\cite{Bruno2012}). Using a dual-symmetric ($w=1/2$) detector, one measures the helicity-resolved Husimi functions $Q^+ (\hat{\mathbf{n}})=Q^{L} (\hat{\mathbf{n}})$ for $(+)$-helicity and 
$Q^- (\hat{\mathbf{n}})=Q^{R} (\hat{\mathbf{n}})$ for  $(-)$-helicity, and identifies their pairs of Majorana stars ($\hat{\mathbf{a}}^+_1,\ \hat{\mathbf{a}}^+_2$) for $(+)$-helicity and ($\hat{\mathbf{a}}^-_1,\ \hat{\mathbf{a}}^-_2$) for $(-)$-helicity. The final expressions then read
\begin{eqnarray}
\mathbf{R}^{\pm\star}\cdot\mathbf{R}^\pm &=& \frac{3}{4\pi} \int d^2 \hat{\mathbf{n}} \ Q^\pm (\hat{\mathbf{n}}) , \\
\mathbf{R}^{+\star}\cdot\mathbf{R}^+ + \mathbf{R}^{-\star}\cdot\mathbf{R}^- &=& 4\, I_{1/2} [A] 
\end{eqnarray}
and
\begin{equation}
\frac{ \text{Im}\left( \mathbf{R}^{\pm\star}\times\mathbf{R}^\pm    \right)}{ 2 \mathbf{R}^{\pm\star}\cdot\mathbf{R}^\pm} = 
\frac{-2 \left(\hat{\mathbf{a}} ^\pm_1 + \hat{\mathbf{a}} ^\pm_2\right)}{3+\hat{\mathbf{a}} ^\pm_1 \cdot \hat{\mathbf{a}} , ^\pm_2} ,
\end{equation}
from which all the physical quantities above can be obtained.


\begin{thebibliography}{99}
\bibitem{sign convention} The sign convention adopted here is such that the physical value of the wave amplitude is $\mathcal{A}(t)= \text{Re}(A e^{-i\,\omega t})$.
\bibitem{gedanken} The process described here assumes that we are able to perform at will such interferometric measurements; that this is not always feasible in practice (in particular for GWs!) is not essential here, as the purpose of this paper is to illustrate concepts, and for such purpose, a \emph{gedankenexperiment} is perfectly satisfactory.
\bibitem{phase definition} The phase reference for the circular-polarization wave is chosen as follows: let us select arbitrarily one point on the unit sphere, that we call the North pole. The phase of $C^L_{\hat{\mathbf{n}}}$ is such that the electric field $\mathbf{E}$ (for electromagnetic radiation) is along the local meridian pointing towards the south, at $t=0$. For gravitational radiation, the phase of $C^L_{\hat{\mathbf{n}}}$ is such that the (gravito-electric) tidal tensor $E_{ij}$ generates a tensile stress along the meridian direction, and a compressive stress along the direction of the parallel, at $t=0$.
\bibitem{footnote-ellipse} This follows directly from the observation that for a given elliptically polarized propagative wave, the orthogonal polarization mode propagating in the same direction has its polarization ellipse rotated by $\pi/2$ around the propagation axis, with a reversed circulation direction.
\end{thebibliography}
\end{document}